\documentclass[12pt]{iopart}
\usepackage{epsfig}  
\usepackage{caption}
\begin{document}

\def\NZ{{\bar N_0}}
\def\NJ{N_{\J}}
\def\J{J/\psi}

\title{Heavy flavor probes of quark matter} 

\author{R. L. Thews  
}

\address{Department of Physics, University of Arizona,
Tucson, AZ 85721 USA}



\begin{abstract}
A brief survey of the role of heavy flavors as a probe of 
the state of matter produced by high energy heavy ion collisions is
presented.  Specific examples include energy loss, initial state gluon saturation, thermalization and flow.  The formation of quarkonium bound states
from interactions in which multiple heavy quark-antiquark pairs are
initially produced is examined in general.  Results from statistical
hadronization and kinetic models are summarized.  New predictions from 
the kinetic model for $\J$ at RHIC are presented. 

\end{abstract}




\section{Introduction}

The references on the web which respond to the search for 
the term ``heavy flavor" are
predominantly physics-related.  However, there are many from chemistry and
food technology.  One of particular interest cites experimental evidence
that the molecular weight of inorganic salts determine the intensity,
or equivalently the amount, of a specific flavor such as saltiness or bitterness.
For the analogous statement applied to heavy quarks we must treat 
the amount of flavor
in a discrete manner (charm, bottom, top) in order to retain the correlation of
``amount" of flavor with the quark mass.

The property of heavy flavors which provides qualitatively different
probes for heavy ion collisions is just that:  heaviness.
The relevant quantities are the current quark masses, generated 
by electroweak symmetry breaking.  
The distance scales corresponding to the inverse of the charm 
mass (1.5 GeV, 0.07 fm)
and bottom mass (5.0 GeV, 0.02 fm) are small compared with the size
of the 
interaction regions in heavy ion collisions and also the size of bound states
in vacuum.    

The following sections survey some characteristic features of heavy 
quark sensitivity to initial state
properties in heavy ion collisions, possible interaction with light
quarks and gluons in probing the spacetime development of a dense expanding
region, and modification of final state hadrons containing heavy quarks
due to interactions with 
a color deconfined medium.  In addition, the utility of heavy quarkonium
states as a signal of deconfinement is updated to include effects of
multiple heavy quark pair production at collider energies.

\section{Initial production of heavy flavor and gluon saturation}

Heavy flavor is not present (at least to any great extent) in the initial
colliding nuclei, so it must be produced initially in the nuclear 
collision itself.  
Production of heavy flavor is one of the so-called "hard probes", since
the mass scale $M_Q >> \Lambda_{QCD}$.  This enables
one to factor out all lower-scale effects and
perform a perturbative expansion within QCD.
The large mass also ensures that this production takes place
at early times, allowing the heavy flavor to probe the subsequent medium
during its entire history.  

The primary production mechanism is through
gluon-gluon interactions, so the total amount is a probe of the initial
state gluon distribution.  
However, nuclear collisions bring
in additional effects, such as shadowing in structure functions, initial
state $k_t$ broadening, and perhaps saturation effects at low-x.
The formation process is expected to scale with the number of
binary nucleon-nucleon collisions $N_{coll}$.  In terms of the number of
nucleon participants, which is taken as a measure of the centrality
of the collisions, the number of binary collisions is expected
to grow as $N_{part}^{4/3}$.

Initial state gluon saturation is a consequence of
very large gluon occupation numbers at low-x.  It is
characterized by yet another scale, the saturation energy scale $Q_s(x)$.
In the color glass condensate picture, $Q_s$ is the scale below which
the $k_t^{-2}$ behavior of the gluon transverse momentum spectra saturates.
In nuclei, one expects an enhancement factor in $Q_s$, proportional
to $A^{1/6}$.
If the saturation scale becomes greater than the other scale in the
problem, the heavy flavor mass, then one may expect collinear factorization
to fail, since the relevant $k_t$ values will not be small compared with $M_Q$.
One consequence is that the heavy quark production process would
scale as $N_{part}$ rather than $N_{coll}$.  However, there may be
residual effects from the heavy quark mass scale even in the region where 
$N_{coll}$ scaling holds for the total rate. An example of this
effect appears in a calculation of the hardening of the D meson spectrum
predicted for central collisions at RHIC\cite{tuchin}.

\section{Thermal production}

There may be subsequent production of heavy flavor from collisions of
secondary partons in the medium, the amount of which is still subject to
considerable theoretical uncertainty.  
However any thermal production will be very sensitive to the ratio
$M_Q/T$, and hence may provide information on the temperature of
an equilibrated medium.

\section{Interaction with a hadronic or deconfined medium}

The utility of heavy flavor as an independent  probe of the dense 
medium relies on
the mass dependence of energy loss/quenching. 
A first order effect is due to the dead cone, which sets the minimum
angle of radiative gluon emission relative to the heavy quark 
direction at $M_Q$/E \cite{deadcone}.  In addition, there will be a
change in the gluon formation time which reduces the coherence 
length in the medium and also the LPM effect\cite{wang}.  Several groups 
have recently calculated the resulting effect on the 
medium-induced radiation from a heavy quark. All conclude that some 
``filling in" of the dead cone will occur \cite{kampfer,gyulassy,wiedemann},
but there will be a residual qualitative difference between
energy loss of heavy relative to light flavors.  This effect may
be observable in the quenching of flavor-tagged jets in AA interactions.
A study of the effects expected at LHC energy is underway\cite{dainese}.
The nuclear suppression ratios to be measured appear to be sensitive to flavor
content under various assumptions of energy loss behavior.

\section{Thermalization and flow}
Finally there is the question of possible thermalization and flow of
heavy flavor in the medium.  
This would require a large opacity of the medium, perhaps the sQGP
fluid.  One would then anticipate a reduced energy loss for the
heavy quarks, since they would be effectively at rest with respect
to the dense partonic medium. At present, one can only infer the production
of heavy flavor hadrons in AA collisions at RHIC 
(in this case charm) by measuring the 
decay leptons at high $p_t$.  However, it has been shown\cite{nagle}
 that this
lepton spectrum for $p_t$ less than about 2.5 GeV is essentially
independent of the initial charm quark momentum distribution.
Elliptic flow is, however, a more sensitive probe \cite{molnar}.
Recent measurements of elliptic flow
for high-$p_t$ leptons provide a hint that the parent heavy quarks may 
be providing the underlying flow\cite{flow}. 

\section{Quarkonium formation from uncorrelated heavy quarks}

Quarkonium formation dynamics in heavy ion collisions originated the interest in
heavy flavor, in particular the $\J$, as a probe of color 
deconfinement.  This of course was initiated by the seminal idea of
Matsui and Satz\cite{matsuisatz}, who noted that plasma screening of the color
force would ``melt" the $\J$. During the period of deconfinement
the c and $\bar{c}$ quarks would generally diffuse apart. 
At hadronization they
would have a much greater probability of combining with a light
quark to form open charm mesons, hence the $\J$ suppression, {\it provided
that no other source of charm quarks would be present}.

An additional part of the scenario was added by Kharzeev and Satz\cite
{dimasatz}, who noted
that ``ionization" of $\J$ in medium would be  readily accomplished by
deconfined thermal gluons.  
The corresponding process using 
gluons confined within light flavor hadrons was shown not to 
produce any significant suppression, due to the different
gluon momentum spectrum in bound states.  

The NA50\cite{cortese}
experiment at the SPS have shown the existence of such a
suppression, with respect to an ``expected" baseline due to
absorption by collisions with nucleons.  
Extrapolation of the suppression scenario
to RHIC energy resulted in a variety of predictions.  The
initial $\J$ data from PHENIX\cite{phenixjpsi} did not have sufficient
statistics to differentiate among 
these predictions\cite{bass}, and the upcoming results from Run 4 
are eagerly awaited.

The primary topic remaining is an investigation of
the effects of high energy (RHIC and above) nuclear collisions, in
which one expects that multiple quark-antiquark pairs will be produced in
an individual nucleus-nucleus collision.  
It is clear that this 
must happen at some point as energy increases, and there is some
initial data on open charm production at RHIC with which to compare.
The basic premise of work in this environment is that one can
probe color deconfinement in a qualitatively different way,
due to the presence of multiple pairs of heavy quarks.
In this situation one avoids the
Matsui-Satz condition of no additional heavy quarks with which
to recombine.  
Then there should be a contribution to heavy quarkonium formation
which utilizes combinations of initially uncorrelated
quark and antiquark, leading to a {\it quadratic} increase with
the total number of heavy flavor quarks in the event.  
There are two specific models
which implement this scenario.  

\subsection{Kinetic formation model}
The kinetic model of quarkonium formation in a region of
color deconfinement was first formulated\cite{bc} to estimate a
possible enhancement of the rate
for $B_c$ production in heavy ion collisions relative to that in
single nucleon-nucleon interactions. The subsequent application to
$\J$ formation\cite{jpsiform} found a more significant effect, due
to the formation rate quadratic dependence on total charm quark
numbers.  The physical picture in this case is quite simple.  The final
$\J$ population 
is determined by a competition between the rate of 
dissociation via gluon-initiated ionization, and the formation
process which is simply the inverse of dissociation.   Recent lattice
calculations of spectral functions\cite{karsch, hatsuda} which indicate
that $\J$ states survive in a medium substantially above
the deconfinement temperature support this dynamical picture.  The centrality
and time dependence of the deconfinement region was modeled according
to an isentropic expansion with initial temperature as a parameter, and
a spatial dependence determined by a Glauber model of participant
density in the transverse plane.   Initial results indicated that
an enhancement of $\J$ production at RHIC (relative to scaling of
binary nucleon-nucleon collisions) could occur.  However, the magnitude
of this effect was found to be very sensitive to the momentum spectrum 
of the participating charm quarks, in addition to the overall quadratic
dependence on initial charm quark production.  The initial calculations 
used Gaussian transverse momentum distributions in a flat rapidity
interval of varying width.  Subsequent calculations used more realistic
distributions taken from pQCD results\cite{thewssqm2003}, which
allowed the identification of a restricted region of 
model parameter space constrained
by the initial PHENIX results.  Owing to the large experimental uncertainties,
however, these restrictions do not allow either a confirmation or
refutation of the basic formation process.  

However, the model also makes predictions for the transverse momentum
and rapidity spectra of $\J$ which are formed in medium.  We would expect
in general that these spectra will differ from initial production or 
statistical hadronization.  For this calculation, we have generated
a sample of $c\bar c$ pairs using pQCD at NLO for RHIC energy\cite{mlm},  
which are
used to define the initial quark momentum distribution.  All possible
pairs of  
one c and one $\bar c$ are then weighted by a formation cross section
to generate a set of formation events.  For purposes of identification, we
denote the pairs in which the c and $\bar c$ originated from the same
initial pair as ``diagonal" and all other pairs as ``off-diagonal". 
We use the ``OPE-motivated" cross section, which can be viewed as the QCD
analog of atomic photodissociation. The inverse process then provides
the cross section for quarkonium formation. 
These cross sections
must be supplemented by the differential dependence, which we
adapt from the matrix element appropriate for the coupling of the bound state 
color dipole with gluons.  The resulting set of weighted differential 
formation probabilities can then be used to calculate the spectra of
formed $\J$.

\begin{equation}
{{dN_{\J}} \over d^3 P_{\J}} = \int{{dt} \over {V(t)}}
\sum_{i=1}^{N_c} \sum_{j=1}^{N_{\bar c}} {\it {v}_{rel}} 
{{d \sigma} \over d^3 P_{\J}}(P_i + P_j \rightarrow P_{\J} + X)
\end{equation}
Note that the formation magnitude exhibits the explicit quadratic
dependence on total charm, normalized by the inverse of the 
system volume.

\begin{figure}[htb]
\begin{minipage}[t]{80mm}
\epsfig{clip=,width=7.5cm,figure=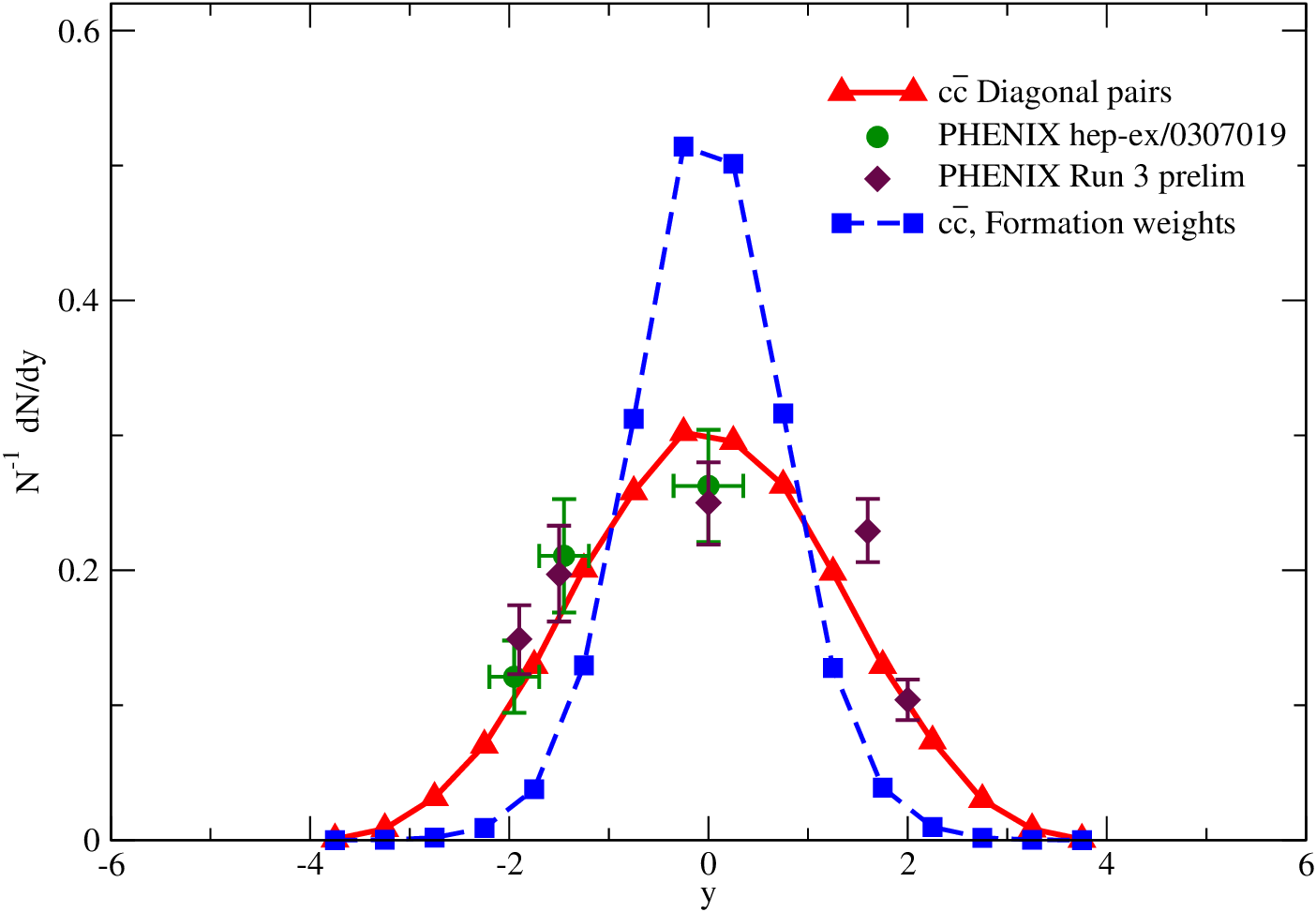}
\caption{\small \\Rapidity distribution for diagonal c$\bar c$ pairs.}
\label{ppyspectra}
\end{minipage}
\hspace{\fill}
\begin{minipage}[t]{75mm}
\epsfig{clip=,width=7.5cm,figure=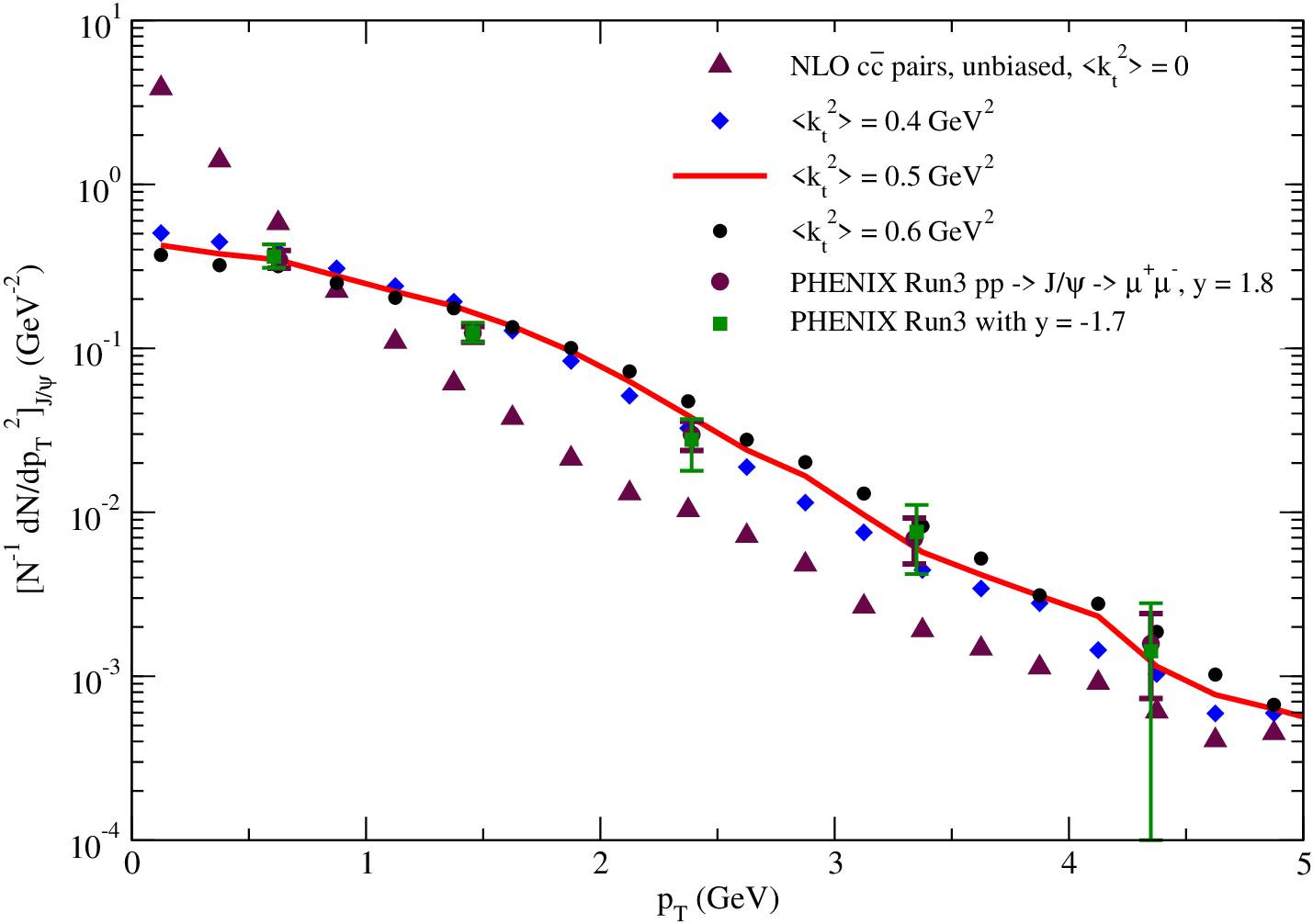}
\caption{\small Transverse momentum distribution for diagonal
c$\bar c$ pairs.}
\label{ppptspectra}
\end{minipage}
\end{figure}

As a baseline test, we first compare with PHENIX data for $\J$
rapidity and transverse momentum in pp interactions\cite{phenixpp}.
We use only the diagonal pairs as appropriate for the pp case.  To
emphasize the formation kinematics, we will concentrate on the
normalized spectra.   
Figure \ref{ppyspectra} shows the rapidity spectra.  One sees that the
diagonal $c\bar c$ spectrum (triangles) agrees quite well with the data. 
For comparison, we show the effect of weighting the pairs with 
the formation probability in a color-deconfined scenario (squares), 
which is inconsistent with the pp data.  This confirms our expectations 
that
the formation model is not applicable in pp interactions.  

\begin{figure}[htb]
\begin{minipage}[t]{80mm}
\epsfig{clip=,width=7.5cm,figure=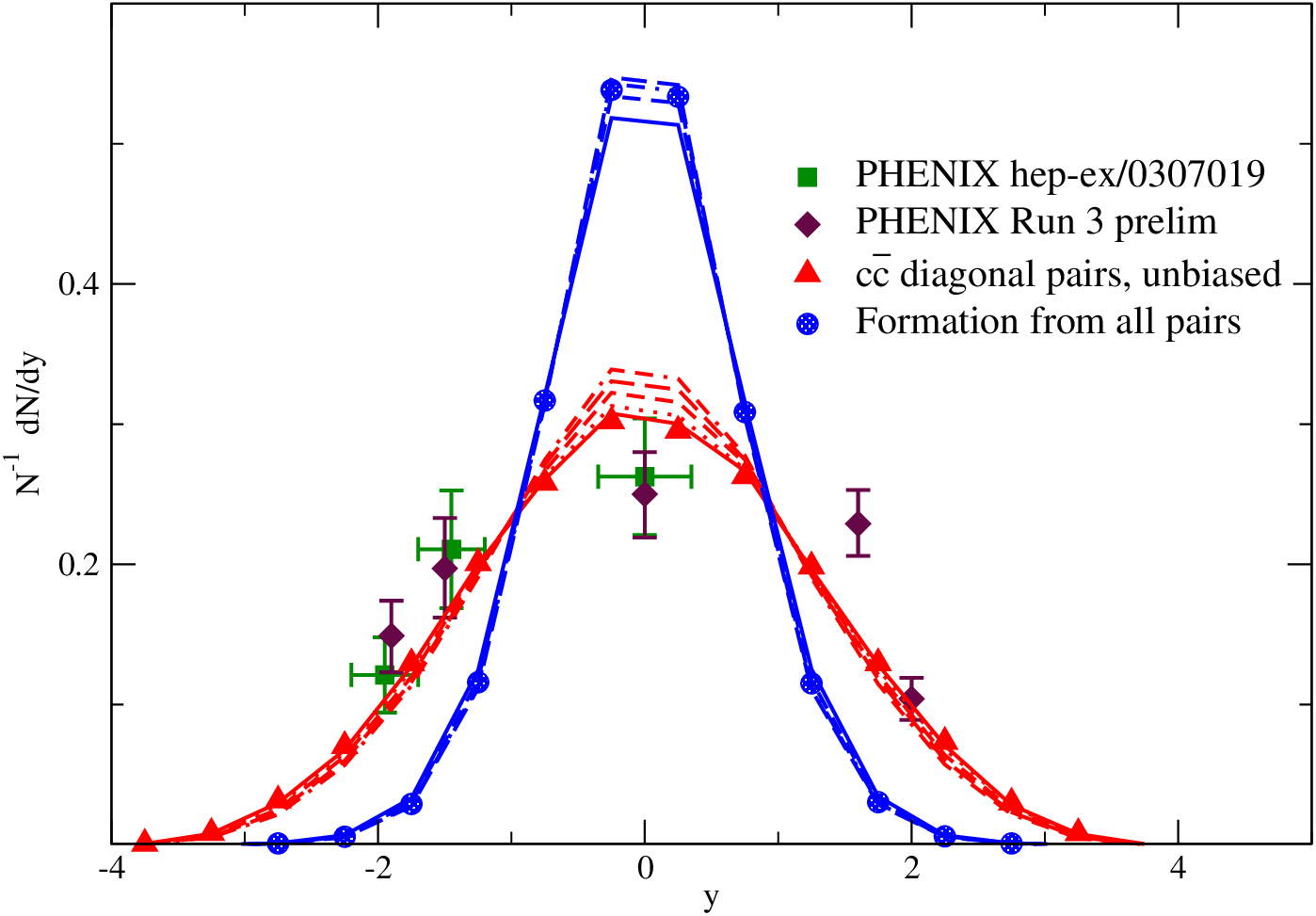}
\caption{\small Prediction for $\J$ rapidity distribution from
formation process in Au-Au collisions at RHIC.}
\label{ypredictions}
\end{minipage}
\hspace{\fill}
\begin{minipage}[t]{75mm}
\epsfig{clip=,width=7.5cm,figure=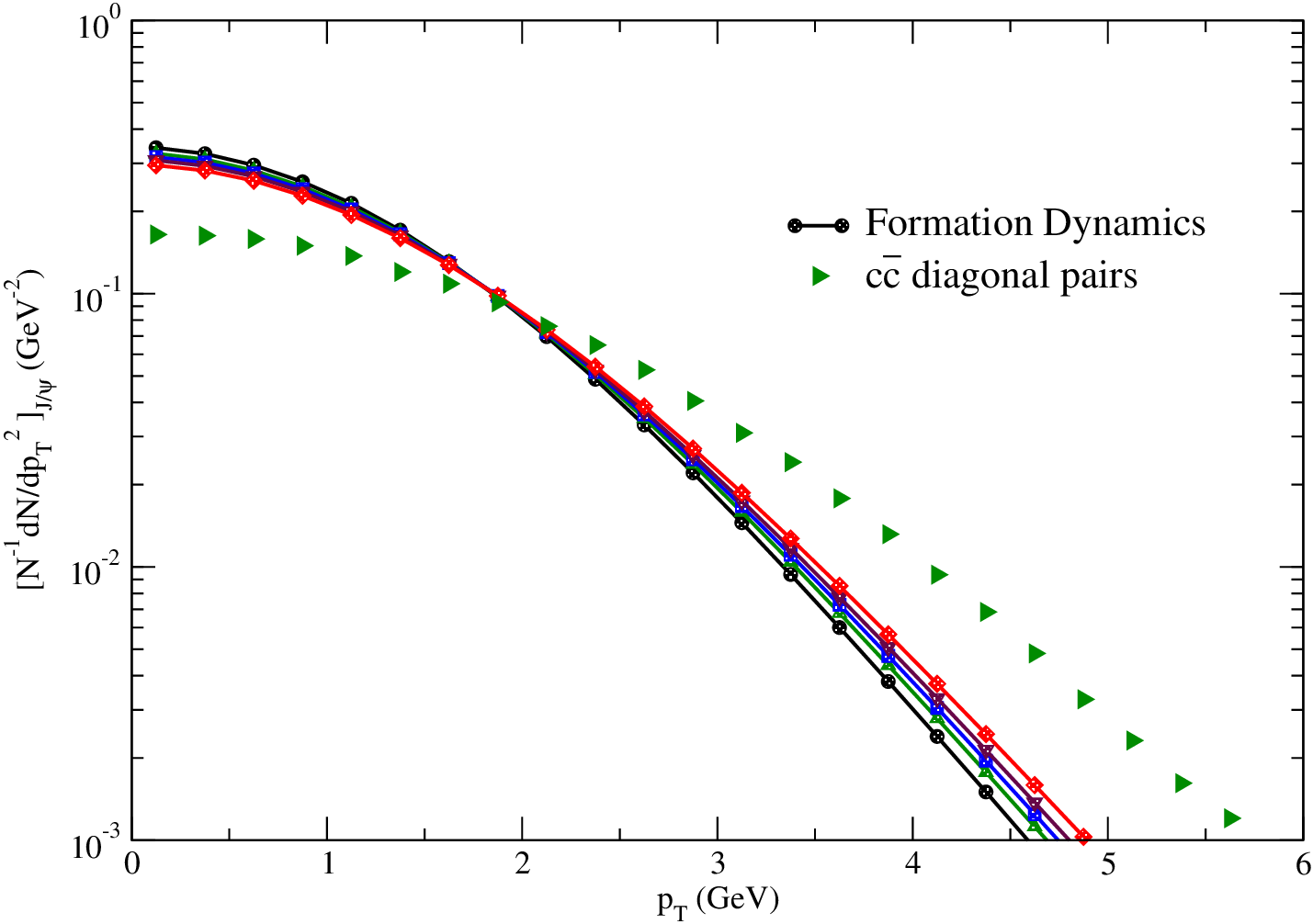}
\caption{\small Prediction for $\J$ $p_t$ distribution
from formation process in Au-Au collisions at RHIC.}
\label{ptpredictions}
\end{minipage}
\end{figure}

Figure \ref{ppptspectra}
shows the corresponding transverse momentum spectra.
The set of curves from the unbiased diagonal $c \bar c$ pairs result from
augmenting the quark initial momenta with a transverse momentum 
``kick" to simulate confinement and initial state effects.  One sees that
the data restricts the magnitude of this kick, parameterized by
a Gaussian distribution with $<k_t^2>_{pp} = 0.5 \pm 0.1 GeV^2$.
To extend this to formation in Au-Au collisions, we must extract
the appropriate $k_t$ for initial state effects in the nucleus.
We use PHENIX data for $\J$ in d-Au collisions, which shows that
the $p_t$ spectra are broadened relative to that in pp interactions\cite
{phenixdau}.  This results in an estimate for 
$<k_t^2>_{Au-Au} = 1.3 \pm 0.3 GeV^2$, which was
utilized in the formation calculations in Au-Au interactions.
Figure \ref{ypredictions} shows rapidity spectra for $\J$
formation utilizing both diagonal and off-diagonal
pairs.  For comparison the data and diagonal pair curves in pp 
interactions are also shown.  One sees that the $k_t$ kick has
virtually no effect, and that the formation mechanism predicts 
a narrowing of the rapidity distribution compared to pp interactions.

Figure \ref{ptpredictions} shows the corresponding transverse momentum
spectra for the allowed range of $k_t$ in Au-Au collisions.  
For comparison we show the
distribution utilizing only unbiased diagonal $c \bar c$ pairs with 
$<k_t^2>$ = 1.3 GeV$^2$, which should be relevant if all of the
$\J$ were produced directly from the initial pairs.  Again the
prediction of this formation mechanism is for a narrowing of
the transverse momentum distribution.
(Of course, both of these distributions would be modified by
the competing dissociation process during the expansion phase,
but one would anticipate a similar effect on each which would
preserve the relative comparison.)

Finally, we contrast these formation transverse momentum predictions
with results which follow if the initially-produced quarks 
were to thermalize and flow with the medium, and subsequently form
$\J$ via the kinetic model.  
For these calculations we used a
``blast wave" parameterization for the charm quark transverse momentum
distribution and a flat rapidity plateau over four units.

\begin{equation}
{{dN_{c,\bar c}} \over dp_t^2} \propto m_t \int_0^R r dr
I_0[{p_t \over T} sinh({r \over R} y_t^{max})]
K_1[{m_t \over T} cosh({r \over R} y_t^{max})]
\end{equation}

\begin{figure}[htb]
\begin{minipage}[t]{80mm}
\epsfig{clip=,width=7.5cm,figure=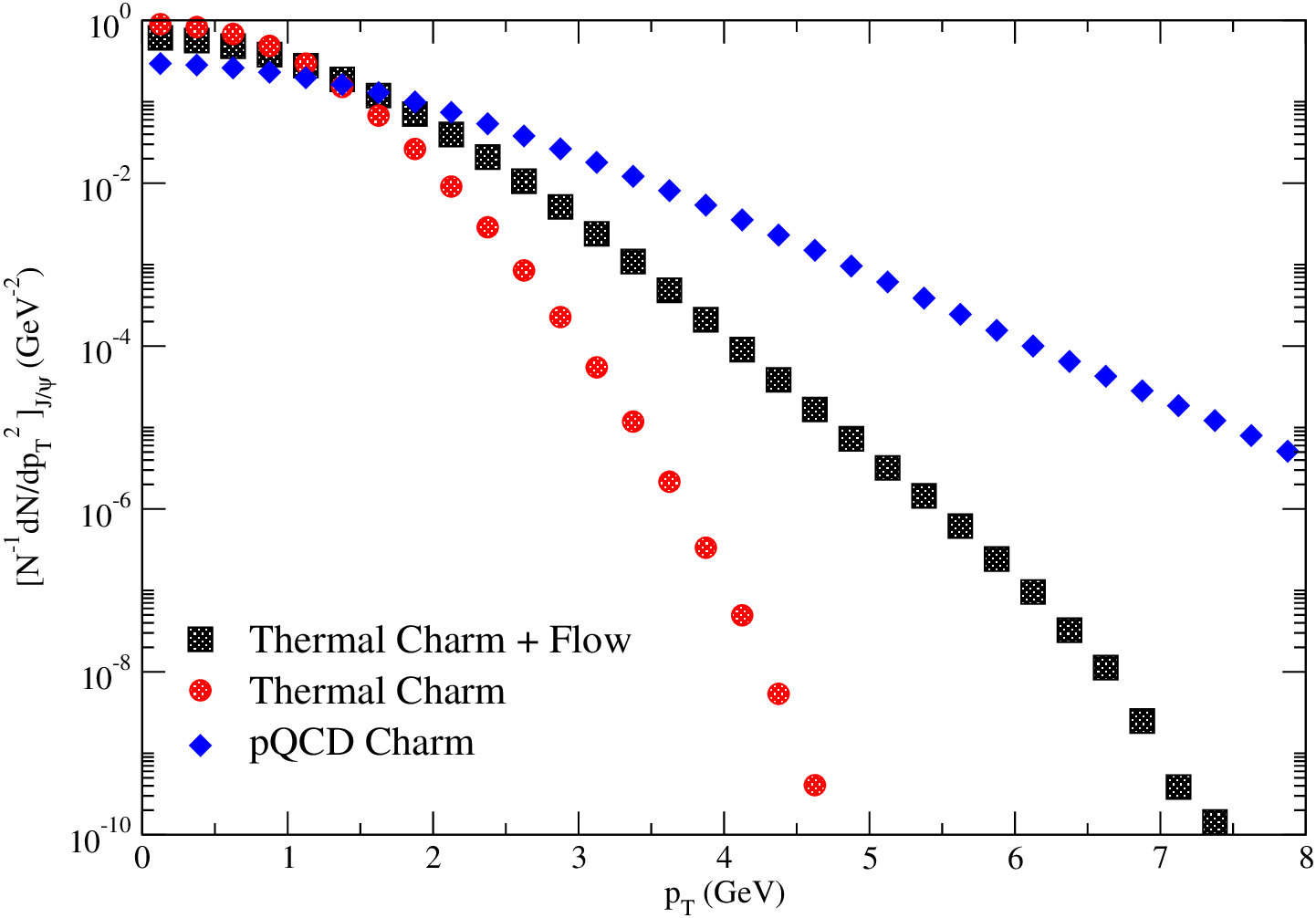}
\caption{\small Comparison of $\J$ formation $p_t$ spectra for different
charm quark distributions.}
\label{blast}
\end{minipage}
\hspace{\fill}
\begin{minipage}[t]{75mm}
\epsfig{clip=,width=7.0cm,figure=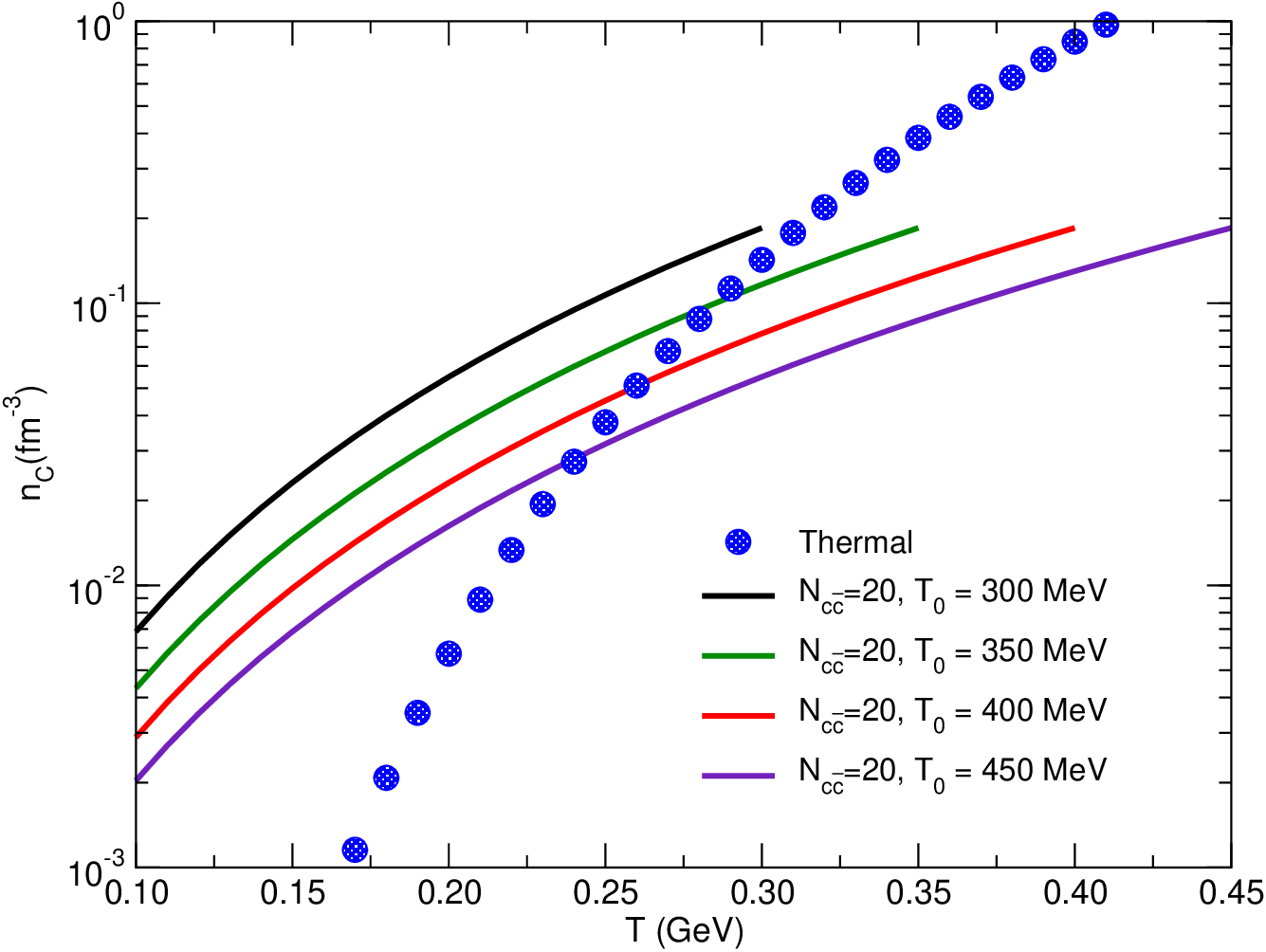}
\caption{\small Comparison of charm quark thermal density with
$N_{c\bar c}$=20 in expanding plasma.}
\label{thermaldensity}
\end{minipage}
\end{figure}

The parameter $y_t^{max}$ specifies the maximum rapidity for the 
linear transverse expansion profile.
In Figure \ref{blast} we show the results for thermal charm quarks with
T = 128 MeV, and for thermal plus flow ($y_t^{max}$ = 0.65).  Shown
for comparison is formation with pQCD charm quarks.  One sees that
the effects of thermal distributions with or without flow for the
charm quarks would be readily distinguishable from the situation
in which pQCD charm quarks participate directly in the formation
process without further interaction in the medium.

\subsection{Statistical hadronization model}
The initial formulation of the statistical hadronization
model for multiple charm pairs in nuclear collisions was
done by Braun-Munzinger and Stachel\cite{pbms}. 
It is clear that charm density exceeds chemical equilibrium
density for T less than $T_c$ for typical volumes and temperatures
appropriate for RHIC conditions.  This is shown in Figure \ref{thermaldensity}.
 One sees that the decrease in density due to expansion (solid curves) is much
less dramatic than the corresponding decrease in thermal equilibrium
density due to the temperature decrease (solid circles).

The extension of the model to accommodate heavy flavor hadrons is
accomplished by introducing a nonequilibrium parameter $\gamma_c$ which
is determined by conservation of total charm and anticharm.  Since
the thermal density is dominated by the open charm states due to their
lower mass, the factor $\gamma_c$ is directly proportional to
the number of initially-produced c$\bar c$ pairs.  Then the
{\it square } of $\gamma_c$ modifies the thermal equilibrium
density for hidden charm mesons, producing the expected quadratic
dependence on $N_{c\bar c}$.  To get the absolute hadron species yields
one must input a volume, which is usually parameterized as the ratio
of total yield to thermal density of charged hadrons.  The final 
expression is 

\begin{equation}
N_{\J} = 4{{n_{ch}^{therm}n_{\J}^{therm}}\over (n_{open}^{therm})^2}
{N_{c \bar c}^2 \over N_{ch}}.
\end{equation}

For application to collider experiments which do not have full
kinematic coverage, one must extend the formalism to rapidity
densities.  This necessitates additional input, in particular the 
magnitude and centrality dependence of the hadronization volume 
for rapidity density.  Two groups\cite{redlich,kostyuk}
 have produced results which 
can be compared with the initial PHENIX data on $\J$ production
in Au-Au at 200 GeV\cite{phenixjpsi}. Both are in general agreement
with the one data point and one upper limit for two centrality
regions, within the statistics-limited experimental uncertainties.
However, two different assumptions were used by the separate groups
for the total charm production cross section, varying from 390 to
650 $\mu$b.  Since these numbers enter the predictions quadratically,
one sees that there must be substantial flexibility remaining in
these model predictions.   Again, it is anticipated that the
upcoming experimental results will place much more severe constraints
on models of this type.    
\ack
This work was supported  by a grant from
the U.S. Department of Energy,  DE-FG02-04ER41318.

\end{document}